\documentstyle[11pt,fleqn]{article}  
\input{psfig.tex}                
\title{Resonant States of Wave Propagation \\ in Disordered System of Bosonic Particles}
\author{A. Kwang-Hua Chu \cite{CHU:2004}}  
\date{P.O. Box 30-15, (Xu-Jia-Hui Post Office), Shanghai 200030, PR China}
\begin{document}      
\maketitle                 
\begin{abstract}           
We demonstrate the effects of an induced disorder (or
a free-orientation : $\theta$ which is related to the relative
direction of scattering of particles w.r.t. to the normal of the
propagating plane-wave front) upon the possible resonance  of the
plane (sound) wave propagating in  Bose gases  by using the
quantum kinetic equations.  We firstly present the diverse
dispersion relations  obtained by the relevant Pauli-blocking
parameter $B$ (which describes the Bose particles when $B$ is
positive) and the free-orientation $\theta$ and then, based on the
acoustic analog, address the possible resonant states.

\vspace{2mm} \noindent PACS : \hspace*{2mm}   42.25.Bs; 02.30.Jr;
05.30.Jp; 34.50.-s; 42.50.Ar; 42.50.Gy; 43.35.Ae; 43.35.Gk;
67.40.Mj \newline
\end{abstract}
\doublerulesep=6mm    
\baselineskip=6mm  
\bibliographystyle{plain}
\section{Introduction}
Bose-Einstein condensation, as is known, occurs when a macroscopic number of bosons
piles down to the same lowest single-particle state.
In the presence of a trapping potential, there arises
a whole spectrum of energy levels ($E_i$) associated with the stationary solutions to the
Gross-Pitaevskii equation. Each of these stationary solutions is, by definition, a topological
coherent mode [1-2]. The lowest energy level corresponds to the standard Bose-Einstein
condensate, while the higher states describe various nonground-state condensates. The
latter can be generated by alternating fields, whose frequencies should be in resonance with
the related transition frequencies [2-3]. It is feasible to realize an oscillatory modulation
of either the trapping potential or of the atomic scattering length. The trapping potential
can be either single-well or multi-well.  Meanwhile,
the theoretically predicted low-energy collective
oscillations of the Bose condensate have been
experimentally confirmed by laser imaging techniques [1].
Moreover, the dynamics of the collective oscillations of the condensate has been theoretically
studied beyond the linear regime, showing
strong enhancement of the amplitude dependence of
frequencies in presence of resonances (cf. Pitaevskii and  Stringari in [1]).
\newline
Very recently, optical lattices and atomic
transport therein have attracted  new attention with the
achievement of Bose-Einstein condensation by purely optical means
and the observation of a superfluid to Mott insulator transition
in a gas of ultracold atoms [4]. However, no detailed study of the dependence of
spatial diffusion on the different directions of an anisotropic lattice  have been
performed so far.  In fact, cold atomic samples also
constitute an ideal system for the study of complex nonlinear
phenomena and allows for the direct observation of wave packet
dynamics in real space on a macroscopic scale [1-4]. Above mentioned progresses
are closely relevant to the quantum-mechanic many-body phenomena subjected to
extremely confinements. One special interest is related to the dynamic resonances
in cold Bose gases [2-3].
\newline On the other hand, emerging interests
in the wave propagation in the random, disordered, and granular
media under the influence of spatial confinement as well as
studies of collision phenomena in rather cold gases, e.g., Bose
gases have stimulated intensive researches recently [5-8]. For
instance, Varshni investigated the spectra of helium (bosonic
particles) at high pressures [5]. The properties of helium atoms
confined to move in restricted geometry have been of considerable
interest during the last several decades [9-10]. The confined
helium atom, however, needs much more complicated calculations
than the confined hydrogen atom. Varshni considered the effect of
confinement on some of the lower lying excited states and the
resulting optical spectrum from transitions between these states.
His results could be applicable to (a) bubbles of helium implanted
in a variety of materials (say, metals [11]), the physical state
of helium in these bubbles had to be inferred by indirect means;
(b) high-pressure helium plasmas [12] (to provide a quantitative
diagnostic for plasma density); (c) astrophysics [5]. \newline We
noticed that acoustical analogs [13-14] considering
continuum-mechanic and quantum-mechanic approaches are currently in rapid
progress for both theories and measurements. The energy ($E$) adopted in quantum-mechanic
formulation directly links to the acoustical frequency ($\omega$) considered
in classic (continuum-mechanic) or semi-classic (kinetic) formulation
due to the existence of an acoustical analogy [14]. Classical
systems could be used to study time-dependent potential fields and
nonlinear effects, which are very difficult and time-consuming to
treat numerically or analytically in quantum-mechanic ways.
Motivated by the need to understand the wave dynamics of cold Bose gases [15-16] at the
finite temperature ($T > T_c$) under strong confinement (which is beyond the
hard-core approximation we adopted here although our approach can
provide the theoretical understanding about how the collision
frequency or the rarefaction parameter can tune the wave
dispersion), an investigation for observing the microscopic localization (which
will induce resonances in confined microdomain) [17-20] using the quantum kinetic model
was performed and will be presented here by taking into account the acoustic analog [14].
\newline In this paper, considering the quantum
(discrete) kinetic model and the Uehling-Uhlenbeck collision
term which could describe the collision of a gas of hard-sphere
Bose-particles by tuning a Pauil-blocking parameter $\gamma$ [21]
(via a {\it Pauli blocking factor} of the form $1+\gamma N_0$ with
$N_0$ being a normalized number density giving the number of
particles per cell in phase space), we will investigate the
possible resonant states when plane waves propagates in
(hard-sphere) Bose gases by introducing a disorder (say, induced
by a high pressure or external field) or free-orientation
($\theta$ which is related to the relative direction of scattering
of particles with respect to the normal of the propagating
plane-wave front) and then obtaining the diverse dispersion
relations which can thus be applied to the acoustical analogs
[13-14]). This presentation will provide more clues to the studies of
the quantum wave dynamics in Bose gases under suitable confined
conditions and the possible appearance of the resonant states
which are linked to the particles (number) density, induced
disorder or relative free-orientation ($\theta$) and their energy
states (the analogy between $E$ and $\omega$) [2-3].
\newline
The necessary verification of our approach with the previous
available approaches [22-27] will be checked firstly.
Our preliminary results show that, for the dispersion part (ratio
of the (phase) wave speed to that of hydrodynamical limit), the
qualitative agreement (for hydrodynamic regime) with Cowell {\it
et al.}'s result [26] or Andrews {\it et al.}'s result [22] is
quite good. As for the damping or attenuation part of ours, the
qualitative agreement with Jackson and Zaremba's (m=0 mode) result
[28] looks also rather good. Our results show that,
as illustrated below, the localized or resonant states will strongly depend on the
energy ($E$ or $\omega$), the effective scattering cross-section,
the number density once the disorder and Pauli-blocking parameters
are selected. We also found that, as
$\theta=\pi/4$, $\theta$ being a disorder parameter, there exist
possible resonant states which are similar to those reported in
[13-14,17-19,29].
\section{Formulations}
We firstly make the following assumptions before we introduce the
general equations of our model [14,25] : \newline (1) Consider a
gas of identical particles of unit mass and a shape of a disk of
diameter $d$, then each particle $i$, $i=1,\cdots,N$, is
characterized by the position of its center $q_i$ and its velocity
$v_i$. We also have the geometric limitations : $|q_i -q_j|$ $\ge
d$, $i\not =j$. This is illustrated schematically in Fig. 1.
\newline (2) Each particle moves in the plane with a relative
velocity (in a centre-of-mass coordinate system) belonging to a
discrete set ${\cal V}$ of 4 velocities with only one speed (due to conservation
of momentum and energy) in the
plane (4 possible different directions) during a binary encounter.
The velocity modulus $c$ is a reference speed depending on
the reference frame and specific distribution of particles.
\newline (3) The collisional mechanism is that of rigid spheres,
that is, the particles scatter elastically and they change their
phase states instantaneously, preserving momentum. Only binary
collisions are considered, since multiple collisions here are
negligible.
\newline The collisions between two particles (say $i$ and $j$)
take place when they are located at $q_i$ and $q_j=q_i-d {\bf n}$,
where ${\bf n}$ is the unit vector joining their centers. After
collisions the particles scatter, preserving momentum, in the
directions allowed by the discrete set ${\cal V}$. In other words,
particles change according to
 $(q_i, v_i) \rightarrow (q_i, v^*_i)$, 
 $(q_j, v_j) \rightarrow (q_j, v^*_j)$ .
The collision is uniquely determined if the incoming velocity and
the impact angle $\psi$, $\psi \in$ [$-\pi/2$,$\pi/2$], are known,
which is defined as the angle between $v_i$ and ${\bf n}$ or ${\bf
n}(\psi)$ = ($\cos$ [$\psi+(k-1)\pi/2$], $\sin$
[$\psi+(k-1)\pi/2$]), $k=1, \cdots, 4$.  \newline From the
selected velocities we have two classes of encounters, i.e.
$\langle v_i, v_j \rangle$ = $0$  and $\langle v_i, v_j \rangle$ =
$-c^2$, respectively. \newline (a) In the first class momentum
conservation implies only : encounters at $\pi/2$ with exchange of
velocities 
 $v_i =v^k \rightarrow v_i^* =v^{k+1}$, \hspace*{2mm} $v_j =v^{k+1}
 \rightarrow v_j^* =v^k$, $k=1,\cdots, 4$,  
in the case $\psi \in [ - \pi/2, 0]$, and
 $v_i =v^k \rightarrow v_i^* =v^{k+3}$, \hspace*{2mm} $v_j =v^{k+3}
 \rightarrow v_j^* =v^k$, 
in the case  $\psi \in$ $[0, \pi/2]$.  \newline (b) Similarly,
$\langle v_i, v_j \rangle$ = $-c^2$;  \newline (i) Head-on
encounters with impact angle $\psi=0$ such that
 $v_i =v^k \rightarrow v_i^* =v^{k+2}$, \hspace*{2mm} $v_j =v^{k+2}
 \rightarrow v_j^* =v^k$, $k=1,\cdots, 4$, \newline 
(ii) Head-on encounters with impact angle $\psi \not=0$ such that
 $v_i =v^k \rightarrow v_i^* =v^{k+1}$, \hspace*{2mm} $v_j =v^{k+2}
 \rightarrow v_j^* =v^{k+3}$, \hspace*{2mm} $\mbox{if $\psi \in$ $[-\pi/2,
 0]$}$ ,  
 $v_i =v^k \rightarrow v_i^* =v^{k+3}$, \hspace*{2mm} $v_j =v^{k+2}
 \rightarrow v_j^* =v^{k+1}$, \hspace*{2mm} $\mbox{if $\psi \in$ $[0,
 \pi/2]$}$.  The schematic presentation is shown in Fig 2. 
For grazing collisions, that is $\langle {\bf n}, v_i \rangle$=
$\langle {\bf n}, v_j \rangle$ = $0$, we put $v^*_i =v_i$, $v_j^*
=v_j$.
\newline \setlength{\unitlength}{1.0mm}
\begin{picture}(120,65)(-30,-30)
\thinlines \put(7,-3){\circle{11.0}} \put(12,7){\circle{11.0}}
\thicklines \put(7,-3){\vector(1,-1){15}}
\put(7,-3){\vector(1,2){16}} \put(7,-3){\vector(2,1){15}}
\put(-7,12){\makebox(0,0)[bl]{\large {\bf $v_1$}}}
\put(-9,0){\makebox(0,0)[bl]{\large {\bf $v^*_1$}}}
\put(12,7){\vector(-3,2){15}} \put(12,7){\vector(-2,-1){15}}
\put(24,1){\makebox(0,0)[bl]{\large {\bf $v$}}}
\put(12,-15){\makebox(0,0)[bl]{\large {\bf $v^*$}}}
\thinlines \put(0,-3){\line(4,0){30}} \put(12,7){\line(2,0){15}}
\put(17,22){\makebox(0,0)[bl]{\large {\bf n}}}
\put(-30,-32){\makebox(0,0)[bl]{\small {\bf Fig. 1 \hspace*{2mm}
Schematic diagram of a collision}}}
\end{picture}      
\begin{picture}(120,65)(80,-30)
\thinlines \put(77,-3){\circle{11.0}} \put(82,7){\circle{11.0}}
\thicklines \put(77,-3){\vector(3,-2){15}}
\put(77,-3){\vector(1,2){15}} \put(77,-3){\vector(2,1){15}}
\put(63,12){\makebox(0,0)[bl]{\large {\bf $v^*_1$}}}
\put(61,0){\makebox(0,0)[bl]{\large {\bf $v_1$}}}
\put(82,7){\vector(-3,2){15}} \put(82,7){\vector(-2,-1){15}}
\put(94,1){\makebox(0,0)[bl]{\large {\bf $v_2$}}}
\put(82,-15){\makebox(0,0)[bl]{\large {\bf $v_2^*$}}}
\put(87,22){\makebox(0,0)[bl]{\large {\bf n}}}
\put(60,-32){\makebox(0,0)[bl]{\small {\bf Fig. 2\hspace*{2mm} A
{\it head-on} collision}}}
\end{picture}
\vspace{4mm}

\noindent We then assume that the gas (i.e., only a two-body
encounter is possible) is composed of identical hard-sphere
particles of the same mass [14,25]. The possible velocities of
these (N) particles are restricted to, e.g., : ${\bf u}_1, {\bf u}_2,
\cdots, {\bf u}_p$, $p$ is a finite positive integer ($p\not = $ N).
That is to say, only the velocity space is discretized, the
space and time variables are still continuous. The discrete
number density of particles are denoted by $N_i ({\bf x},t)$
associated with the velocity ${\bf u}_i$ at point ${\bf x}$ and
time $t$. If only nonlinear binary collisions and the evolution of
$N_i$ are considered, we have
\begin{equation}
 \frac{\partial N_i}{\partial t}+ {\bf u}_i \cdot \nabla N_i
 = F_i \equiv \sum^p_{j=1} \sum_{(k,l)} (A^{ij}_{kl} N_k N_l - A^{kl}_{ij}
 N_i N_j),  \hspace*{3mm} i=1,\cdots, p,
\end{equation}
where $(i,j)$ and $(k,l)$ are admissible sets of collisions
[14,21,25],
$i,j,k,l \in$ $\Lambda$ =$\{1,\cdots,p\}$, and the summation is
taken over all $j,k,l$, where $A_{kl}^{ij}$ are nonnegative
constants (related to transitional rates) satisfying
  $A_{kl}^{ji}=A_{kl}^{ij}=A_{lk}^{ij}$,  
 $A_{kl}^{ij}$ $({\bf u}_i +{\bf u}_j -{\bf u}_k -{\bf u}_l )=0$,
 $A_{kl}^{ij}=A_{ij}^{kl}$ [14,25].  
$F_i$ is the (discrete) approximation of the  collisional integral in the conventional continuous
kinetic theory.
The conditions defined for the discrete velocity above requires
that elastic, binary collisions, such that momentum and energy are
preserved.
\newline The collision
operator is now simply obtained by joining $A_{ij}^{kl}$ to the
corresponding transition probability densities $a_{ij}^{kl}$
through $ A_{ij}^{kl}$ =$ S|{\bf u}_i-{\bf u}_{j}|$ $a_{ij}^{kl}$,
where,
 $a_{ij}^{kl} \ge 0$, $\sum^p_{k=1}  a_{ij}^{kl}=1$,
 $\sum^p_{l=1}  a_{ij}^{kl}=1$,
 $\forall i,j=1,\cdots,p$;
with $S$ being the effective scattering or collisional cross-section [14,21,25].
If all $n$ ($p=2n$) outputs are assumed to be equally probable,
then $a_{ij}^{kl}$=$1/n$ for all $k$ and $l$, otherwise
$a_{ij}^{kl}$= 0.  The term $ S|{\bf u}_i-{\bf u}_{j}| dt$ is the
volume spanned by the particle with ${\bf u}_i$ in the relative
motion w.r.t. the molecule with ${\bf u}_j$ in the time interval
$dt$. Therefore, $ S|{\bf u}_i$ $-{\bf u}_{j}| N_j$ is the number
of $j$-particles involved by the collision in unit time.
Collisions which satisfy the conservation and reversibility
conditions which have been stated above are defined as {\it
admissible collisions} [14,25].\newline
Moreover, all the velocity directions after collisions are assumed
to be equally probable. We note that, the summation of $N_i$
($\sum_i N_i$) : the total discrete number density here is related
to the macroscopic density : $\rho \,(=  m_p \sum_i N_i)$, where
$m_p$ is the mass of the particle [21].
\newline
With the introducing of the Uehling-Uhlenbeck collision term
[21] in equation (1) ($F_i$ being replaced or modified),
\begin{equation}
 F_i =\sum_{j,k,l} A^{ij}_{kl} \,[ N_k N_l (1+\gamma N_i)(1+\gamma N_j)-
 N_i N_j (1+\gamma N_k)(1+\gamma N_l)],
\end{equation}
for $\gamma <0$ (normally, $\gamma=-1$) we obtain a gas of
Fermi-particles; for $\gamma
> 0$ (normally, $\gamma=1$) we obtain a gas of Bose-particles,
and for $\gamma =0$ we obtain equation (1) which is for a gas of
Boltzmann-particles [14,24-25].
\newline Considering binary  collision only, from equation (2),
the model of quantum discrete kinetic equation for Bose gases is
then a system of $2n(=p)$ semilinear partial differential
equations of the hyperbolic type (in two dimensional form):
\begin{displaymath}
 \frac{\partial}{\partial t}N_i +{\bf U}_i \cdot\frac{\partial}{\partial
 {\bf x}} N_i =\frac{c S}{n} \sum_{j=1}^{2n} N_j N_{j+n}(1+\gamma N_{j+1})(1+\gamma N_{j+n+1})-
\end{displaymath}
\begin{equation}
 \hspace*{18mm} 2 c S N_i  N_{i+n} (1+\gamma N_{i+1})(1+\gamma
 N_{i+n+1}),\hspace*{24mm} i=1,\cdots, 2 n,
\end{equation}
where $N_i=N_{i+2n}$ are unknown functions, and ${\bf U}_i$ =$ c
(\cos[\theta+(i-1) \pi/n], \sin[\theta+(i-1)\pi/n])$ are the
particles velocities in the centre-of-mass coordinate system ; $c$
is a reference velocity modulus and the same order of magnitude as
the sound speed in the absence of scatters in [13] or [19],
$\theta$ is the orientation starting from the positive $x-$axis to
the $u_1$ direction and could be thought of as a  disorder induced
by high pressure or external field (schematically shown in Fig.
3).
For example, there are admissible collisions $({\bf U}_1,{\bf
U}_{3}) \longleftrightarrow ({\bf U}_2,{\bf U}_{4})$ as n=2 [25].
\newline Since passage of the sound wave causes a small departure
from an equilibrium state resulting in energy loss owing to
internal friction and heat conduction, we linearize above
equations around a uniform equilibrium state (particles' number
density : $N_0$) by setting $N_i (t,x)$ =$N_0$ $(1+P_i (t,x))$,
where $P_i$ is a small perturbation. The equilibrium state here is
presumed to be the same as in Refs. [21-22]. After some similar
manipulations as mentioned in Refs. [30], with $B=\gamma N_0
>0$ (cf. Chu in [21], $B$ defines the proportional contribution
from the Bose gases; if $\gamma
> 0$, e.g., $\gamma=1$), we then have
\begin{equation}
 [\frac{\partial^2 }{\partial t^2} +c^2
 \cos^2[\theta+\frac{(m-1)\pi}{n}]
 \frac{\partial^2 }{\partial x^2} +4 c S N_0 (1+B) \frac{\partial
 }{\partial t}] D_m= \frac{4 c S N_0 (1+B)}{n} \sum_{k=1}^{n} \frac{\partial
 }{\partial t} D_k  ,
\end{equation}
where $D_m =(P_m +P_{m+n})/2$, $m=1,\cdots,n$, since $D_1 =D_m$
for $1=m$ (mod $2 n)$. \newline 
We start to look for the solutions in the form of plane wave
$D_m$= $a_m$ exp $i (k x- \omega t)$, $(m=1,\cdots,n)$, with
$\omega$=$\omega(k)$ because it is related to the dispersion
relations of quasi one-dimensional  plane waves propagating in
(monatomic) hard-sphere Bose gases. So we have
\begin{equation}
 (1+i h (1+B)-2 \lambda^2 cos^2 [\theta+\frac{(m-1)\pi}{n}]) a_m -\frac{i h (1+B)}{n}
 \sum_{k=1}^n a_k =0  , \hspace*{6mm} m=1,\cdots,n,
\end{equation}
where
\begin{displaymath}
\lambda=k c/(\sqrt{2}\omega),  \hspace*{18mm} h(1+B)=h_b=4 c S N_0
(1+B)/\omega \hspace*{6mm} \propto \hspace*{2mm} 1/K_n,
\end{displaymath}
$h$ is the rarefaction parameter of the gas; $K_n$ is the Knudsen
number which is defined as the ratio of the mean free path of Bose
gases to the wave length of the plane (sound) wave .
\newline Let $a_m$ = ${\cal{C}}/(1+i h_b-2 \lambda^2 \cos^2
[\theta+(m-1)\pi/n])$, where ${\cal{C}}$ is an arbitrary, unknown
constant, since we here only have interest in the eigenvalues of
above relation. The eigenvalue problems for different $2\times
n$-velocity model reduces to
\begin{equation}
  1-\frac{i h_b}{n} \sum^n_{m=1} \frac{1}{1+i h_b -2 \lambda^2
    \cos^2\,[\theta+\frac{(m-1)\pi}{n}]} =0.
\end{equation}
\section{Results and Discussions}
We can resolve  the complex roots ($\lambda=\lambda_r +$ i
$\lambda_i$) from the polynomial equation above
and use the numerical way for direct plots. The roots are the
values for the nondimensionalized dispersion (positive real part;
a  ratio of the sound or phase speed with respect to its continuum
or hydrodynamical limit) and the attenuation or absorption
(positive imaginary part), respectively. $B$ could be related to
the occupation number. We plot the main results into figures 4, 5,
6, and 7, respectively. We firstly review the general
characteristic dispersion relations for Bose gases before we
interpret our present results.
\newline Curves in figures  4 and 5 ($\theta \not =\pi/4$) resemble the
conventional dispersion relations of ultrasound propagating in
hard-sphere Boltzmann-statistic gases [30]. Present results show
that as $B$ or $\theta$ increases, the dispersion ($\lambda_r =
k_r c/(\sqrt{2}\omega)$) will reach the hydrodynamical limit (as
$h \gg 1$ the wave speed is independent on the $S$ or the s-wave
scattering length, this result agrees qualitatively with Cowell
{\it et al.}'s results [26] (for $B>0$) or Bruun and Burnett's
[28] (for $B<0$)) earlier. That is to say, the phase speed of the
plane wave in Bose gases (even for small but fixed $h$) increases
w.r.t. to the continuum conditions ($h \rightarrow \infty$) as the
relevant parameter $B$ increases or $\theta$ increases (up to
$\pi/4$).
\newline  Meanwhile, as illustrated in Fig. 5, there always exist peaks or maximua in
our attenuation results (related to the damping of the propagating
wave). This agrees qualitatively with  that of  Jackson and
Zaremba's result (for m=0 mode) [28]. Here, considering the
Pauli-blocking effect, the maximum or peak absorption (or
attenuation $\lambda_i = k_i c/(\sqrt{2}\omega)$) for all the
rarefaction parameters $h$ keeps the same for all $B$ (say,
$B=0.3$ and $B=0.7$). There are only shifts of the maximum
absorption state (defined as $h_{max}$)  when $B$ increases.  It
seems for the same mean free path ($h \propto$ the inverse of
$K_n$) or mean collision frequency of the hard-sphere gases (i.e.
the same $h$ but $h < h_{max}$) there will be more absorption in
larger $\theta$ cases than those of $\theta=0$ states when the
plane wave propagates.
\newline To apply the acoustic analog, we should now introduce the brief idea [14].
In fact, studies of classical wave mechanical systems have some
important advantages over quantum mechanical wave systems even
there are similarities in-between [13]. In a mesoscopic system,
where the sample size is smaller than the mean free path for
elastic scattering, it is satisfactory for a one-electron model to
solve the time-independent Schr\"{o}dinger equation :
\begin{displaymath}
 -\frac{\hbar^2}{2m} \nabla^2 \psi + V' (\vec{r}) \psi = E \psi
\end{displaymath}
or (after dividing by $-\hbar^2/2m$)
\begin{displaymath}
 \nabla^2 \psi + [q^2 - V (\vec{r})] \psi = 0,
\end{displaymath}
where $q$ is an (energy) eigenvalue parameter, which for the
quantum-mechanic system is $\sqrt{2mE/\hbar^2}$. \newline
Meanwhile, the equation for classical (scalar) waves is
\begin{displaymath}
 \nabla^2 \psi -\frac{1}{c^2} \frac{\partial^2 \psi}{\partial t^2}
 =0
\end{displaymath}
or after applying a Fourier transform in time and contriving a
system where $c$ (the wave speed) varies with position $\vec{r}$
\begin{equation}
 \nabla^2 \psi + [q^2 - V (\vec{r})] \psi = 0,
\end{equation}
here, the eigenvalue parameter $q$ is $\omega/c_0$, where $\omega$
is a natural (or an eigen-) frequency and $c_0$ is a reference
wave speed. Comparing the time dependencies one sees the quantum
and classical relation $E= \hbar \omega$ [14]. Thus, the energy $E$ considered, e.g.,
in Refs. [19-20] for the resonant localization in quantum system corresponds to the
frequency $\omega$ considered here especially for the resonant states [14]
by using the acoustical analogy.\newline
Back to our present figures w.r.t. $h$, as now the parameter of
disorder or free-orientation $\theta$ dominates. We can observe
that, there is a continuous trend as $\theta$ increases toward
$\pi/4 \approx 0.7854$. The dispersion ($\lambda_r$; a relative
measure of the sound or phase speed) keeps increasing while the
attenuation or absorption ($\lambda_i$) keeps decreasing as
$\theta$ increases from $0$. At $\theta=\pi/4$, there is no
attenuation and dispersion, i.e., $\lambda_r=1.0$ and
$\lambda_i=0.0$ [30]. This result also provides a good
verification for the experimental side mentioned in [16-18] as
there is no loss for this particular case ($\theta$ being a
disorder parameter but fixed as $\pi/4$). We note that around the
peak $\lambda_i$ state ($h_{min}$) as shown in Fig. 5, there
exists a trend for the absence of diffusion ($\lambda_i$ starts
decreasing rapidly) [18-20].
\newline  Based on the acoustic analog [14], from the definition of $h$ or
$Kn$, $h = f_{collision}/f_{sound}$, where $f_{sound}$  is related
to the classical frequency $\omega$ as mentioned above (cf.
equation (7)) so that it is relevant to the energy $E$ as defined
for the localization, thus we can estimate the localization length
from those figures which vary with $h$; $f_{collision}$ is the
mean collision frequency of the particles. The localization length
defined in [19] is proportional to the (hydrodynamic) mean free
path $l$ ($l$ also depends on the internal frequency $E$ as shown
in [19]) and, comparing the definition of $h$ here, is thus
related to the inverse of $h$ (or, say, the frequency) we used. We
remind the readers that the temperature ($T$) could be linked to
the mean free path or mean collision frequency under prescribed
conditions [24] and the energy (thus $\omega$) could be related to
the temperature ($T$) with the introdution of the Boltzmann
constant $k_B$. Based on these considerations, the relation for
the localization length versus the frequency extracted from our
results (especially in Fig. 5; the attenuation or absorption
defined here is related to the {\it inverse} measure of, say, one
wave length; the maximum absorption then corresponds to the
minimum localization length in Fig. 5(a) of [19]) is qualitatively
similar to that reported in [19]. This observation is now
illustrated in figure 6 where we schematically define the
localization length as the inverse of the wave absorption :
$1/\lambda_i$ (per unit wave length). There also shows the
exponential decay of the the localization length w.r.t. $h$ or the
inverse of the frequency (a corresponding measure to energy $E$ in
quantum-mechanic sense as already explained before). Thus we can
also obtain similar results which resemble that reported in Fig. 5
(a) in [19]. \newline We note that, as the rarefaction parameter ($h =4 c S N_0/\omega$) is fixed
(cf. figure 6 for the schematic localization length),
from its definition and the correspondence between the energy ($E$) and the
frequency ($\omega$), the product of an effective scattering cross-section ($S$)
(or the s-wave scattering length) and the number density of particles ($N_0$) must be
proportional to
$E$ or $\omega$ once $c$  and the localized state are made specific (the Pauli-blocking
and disorder effects  being excluded). Meanwhile, once $E$ or $\omega$ is fixed
for the same situation mentioned above, the localized state will strongly depend on
the $S$ and $N_0$ as well as $B$ and $\theta$. \newline
People might argue that a nonzero $\theta$
would only make the system anisotropic, but not disordered. We
should remind them that the derivation of present quantum kinetic
approach was based on the binary encounter of a system of Bose
particles. Once the mean free path and the centre of mass
coordinate system were introduced (especially when the effective,
admissible collision and the microreversibility which neglects the
history and the correlations when particles traverse in phase
space [14,21] were presumed) the randomness and disorder will
occur although they are illustrated implicitly.  The results
presented here, in fact, also qualitatively resemble that reported
in [29] where it was shown that when a periodic medium with a gap
(in resulting spectra) is (slightly) randomized (like our
disordered case : $\theta \not =0$), possible localization occurs
in a vicinity of the edges of the gap (like that of $\theta=\pi/4$
here; $\theta=0$ is implicit) [29]. As we only consider plane
waves propagating in a hard-sphere gas, which is a kind of hard
(Neumann) scatters [29], then it is interesting that our results
for the dispersion relation resemble those of Neumann cases
(especially Fig. 9) in [29].
\newline To demonstrate the possible resonant states, we summarize our results in figure
7. Our results show that as $\theta$ increases, the maximum
absorption ($\lambda_i$) will decreases continuously except at
$\theta=\pi/4$ where there is a sudden jump (maximum)! This
unusual absorption (or attenuation) peak is similar to those
observations found in the $T_{\lambda}$ transition (temperature)
for liquid helium (Bose liquids) (cf. Figs. 20, 21 in [31]).
The interesting result is that, the absorption value obtained for
$B=0.5$ (Bose gases) is almost twice of that value for $B=-0.5$
(Fermi gases) at $\theta=\pi/4$. Note that, once there are Cooper
pairings in Fermi gases, we can treat them (atomic pairs) as
bosonic particles although the number density of them might be one
half of the original. Thus, this resonant state might be relevant
to the superfluid phase transition or Cooper pair formation tuned
by the disorder. Meanwhile, once we consider the disorder is
induced by the magnetic field, this sudden jump in Fig. 7
resembles that of resonance reported in [32].
\newline As we know, when a gas containing many identical particles is
confined and cooled, the average momentum can be lowered so far
that the typical de Broglie wavelength is larger than the average
separation between the particles. In this case, the gas is said to
be 'degenerate', meaning that the wave functions of neighboring
particles overlap. Degenerate gases exhibit two dramatically
different types of behavior, depending on whether the identical
particles are bosons (such as photons) or fermions (such as
electrons).  \newline To conclude in brief, our calculations here
are useful to the understanding of waves propagating in
microscopically random, disordered or granular media under strong
confinement [2-3,5-7,9-12] and the dispersion management for
atomic matter waves [16].  The direct relation of our results to
the conventional one of static  localization  is not
straightforward but could be understood qualitatively after the
application of the acoustical-analog (i.e. the necessary transform
from our complex $\lambda$ or real $h$ to the conventional
$\omega$, $E$, $V$ (potential barrier), and the characteristic
lengths (cf. [13-14,19]) : mean free path, wave length, etc.). Our results show that,
as illustrated in figure 6, the resonant states strongly depend on the
energy ($E$ or $\omega$), the effective scattering cross-section ($S$),
the number density ($N_0$) once the disorder and Pauli-blocking parameters
are selected. We
shall study other complicated problems in the future [2-3,33-35].
{\small Acknowledgements.
The author is partially supported by the China Post-Dr. Science
Foundation (Grant No. 1999-17).}

\newpage
\section{Appendix : Derivation of Eqn. (4)}
From Eqn. (3), after the linearization,  we then  have, (say, i=m)
\begin{displaymath}
 \frac{\partial }{\partial t} P_m +{\bf U}_m\, \cdot
 \frac{\partial }{\partial {\bf x}} P_m +2 c S N_0 [(P_m +P_{m+n} )+
 \gamma N_0 (P_m +P_{m+n}+P_{\small \sum})+\cdots]=
\end{displaymath}
\begin{equation}
 \frac{c S N_0}{n} \sum_{k=1}^{2n} [(P_k +P_{k+n}+
 \gamma N_0 (P_k +P_{k+n}+P_{\small \sum})+\cdots],
\end{equation}
here, $m=1,\cdots,2n$, $P_{\small \sum}=0$ for n=2 because of the
restriction for the total perturbations in an equilibrium state
and the remaining terms in both sides are higher order terms
related to $(\gamma N_0)^2$. The linearized version of above
equation (for n=2) is
\begin{equation}
 \frac{\partial }{\partial t} P_m +{\bf U}_m\, \cdot
 \frac{\partial }{\partial {\bf x}} P_m +2 c S N_0 (P_m +P_{m+n} )(1+\gamma N_0)=
 \frac{2 c S N_0}{n} \sum_{k=1}^{2n} P_k (1+\gamma N_0).
\end{equation}
 In these equations after replacing
the index $m$ with $m+n$ and using the identities $P_{m+2n}$ =
$P_m$, then we have
\begin{equation}
 \frac{\partial }{\partial t} P_{m+n} -{\bf U}_m\, \cdot
 \frac{\partial }{\partial {\bf x}} P_{m+n} +2 c S N_0 (P_m +P_{m+n} )(1+\gamma N_0)=
 \frac{2 c S N_0}{n} \sum_{k=1}^{2n} P_k (1+\gamma N_0) .
\end{equation}
Combining above two equations, firstly adding then subtracting,
with $A_m =(P_m + P_{m+n})/2$ and $B_m=(P_m -P_{m+n})/2$, we can
have
\begin{equation}
 \frac{\partial }{\partial t} A_m -c\,
 \cos[\theta+\frac{(m-1)\pi}{n}]
 \frac{\partial }{\partial x} B_m +4 c S N_0 A_m (1+\gamma N_0)=
 \frac{4 c S N_0}{n} \sum_{k=1}^{2n} A_k (1+\gamma N_0) , 
\end{equation}
\begin{equation}
 \frac{\partial }{\partial t} B_m +c\,
 \cos[\theta+\frac{(m-1)\pi}{n}]
 \frac{\partial }{\partial x} A_m =0 , \hspace*{3mm} m=1,\cdots,2n.
\end{equation}
From $P_{m+2 n}=P_m$, and with $A_m =(P_m +P_{m+n})/2$ and
$B_m=(P_m -P_{m+n})/2$, we can have $A_{m+n}= A_m$, $B_{m+n}$
=$-B_m$. \newline After eliminating $B_m$, with $B=\gamma N_0$, we
then have
\begin{displaymath}
 \{\frac{\partial^2 }{\partial t^2} +c^2 \cos^2
 [\theta+\frac{(m-1)\pi}{n}]
 \frac{\partial^2 }{\partial x^2} +4 c S N_0 (1+B) \frac{\partial
 }{\partial t}\} D_m= \frac{4 c S N_0 (1+B)}{q} \sum_{k=1}^{n} \frac{\partial
 }{\partial t} D_k  ,
\end{displaymath}
where $D_m =(P_m +P_{m+n})/2$, $m=1,\cdots,n$, since $D_1 =D_m$
for $1=m$ (mod $2 n)$.

\newpage

\oddsidemargin=3mm

\pagestyle{myheadings}

\topmargin=-18mm

\textwidth=17cm \textheight=26cm
\psfig{file=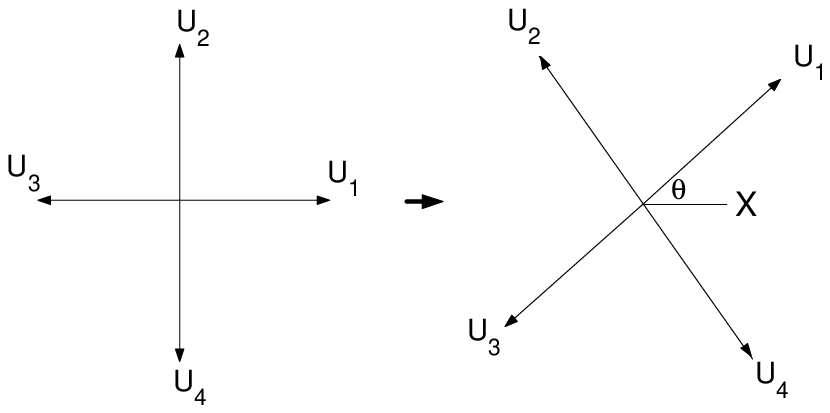,bbllx=0.1cm,bblly=17.5cm,bburx=12cm,bbury=23.8cm,rheight=8cm,rwidth=8cm,clip=}
%
\begin{figure}[h]
\hspace*{6mm} Fig. 3 \hspace*{1mm} Schematic plot for the regular
scattering and the disorder-influenced scattering. \newline
\hspace*{7mm} Plane waves propagate along the $X$-direction.
Binary encounters of $U_1$ and $U_3$ and their \newline
\hspace*{7mm} departures  after head-on collisions ($U_2$ and
$U_4$). Number densities $N_i$ are associated to $U_i$.
\end{figure}

\vspace{8mm}

\psfig{file=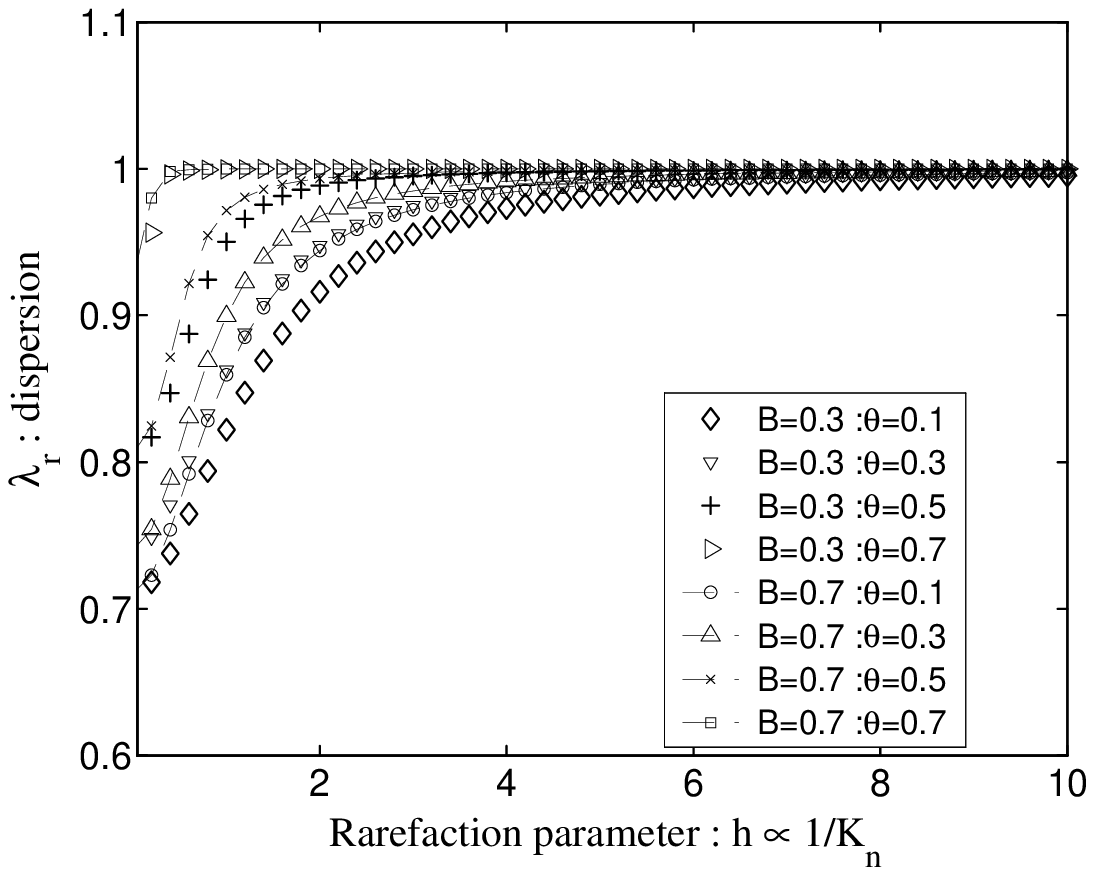,bbllx=0.1cm,bblly=13.5cm,bburx=12cm,bbury=24.2cm,rheight=10cm,rwidth=10cm,clip=}
\vspace{2mm}
\begin{figure}[h]
\hspace*{8mm} Fig. 4 \hspace*{1mm} Disorder or orientational
($\theta$) effects  on the  dispersion ($\lambda_r$). \newline
\hspace*{9mm} $h =4 c S N_0/\omega$, $S$ is the effective
scattering cross section,  $N_0$ is the \newline \hspace*{9mm}
number density, $B=\gamma N_0$ is the Pauli-blocking parameter.
\newline \hspace*{9mm} This result agrees with Cowell {\it et
al.}'s result [26] for the hydrodynamic
\newline \hspace*{9mm} regime ($h \sim O(10)$ here). Wave speed is
independent on $S$ for larger $h$.

\end{figure}

\newpage

\psfig{file=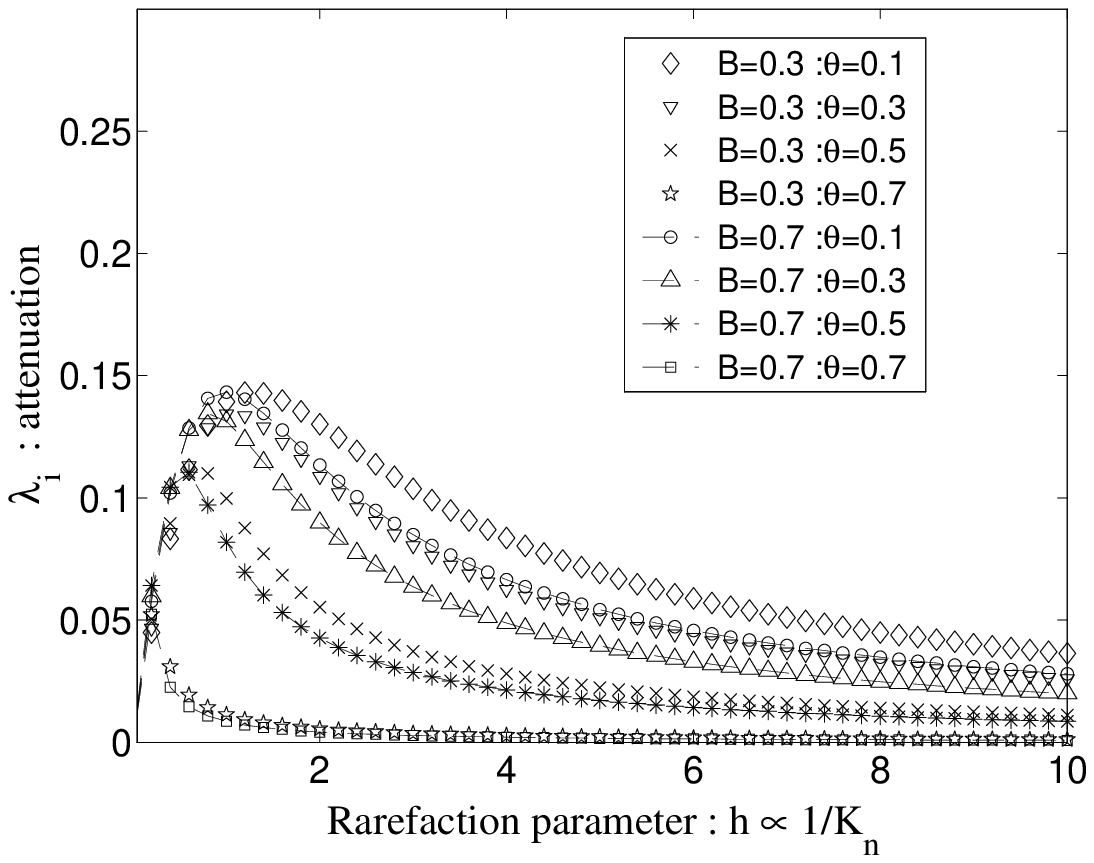,bbllx=0.1cm,bblly=13.6cm,bburx=12cm,bbury=24cm,rheight=9.6cm,rwidth=9.6cm,clip=}
\vspace{2mm}
\begin{figure}[h]
\hspace*{8mm} Fig. 5 \hspace*{1mm} Disorder or orientational
($\theta$) effects  on the attenuation ($\lambda_i$). \newline
\hspace*{9mm} This result agrees with Jackson and Zaremba's result
(m=0 mode) [28]. \newline \hspace*{9mm} There always exists a peak
or maximum $\lambda_i$.

\end{figure}

\vspace{3mm}
\psfig{file=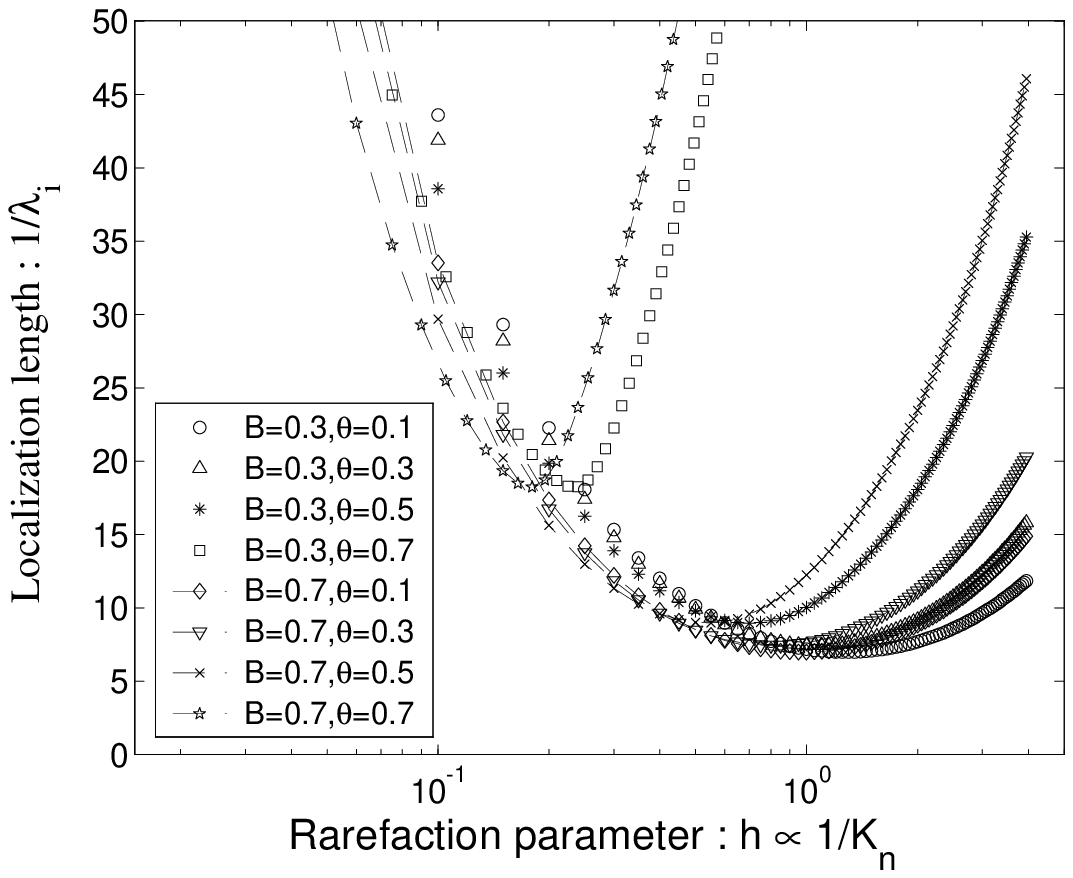,bbllx=0.1cm,bblly=13.6cm,bburx=12cm,bbury=23.8cm,rheight=10cm,rwidth=10cm,clip=}
\vspace{2mm}
\begin{figure}[h]
\hspace*{6mm} Fig. 6 \hspace*{1mm} Disorder ($\theta$) effects on
the localization length ($1/\lambda_i$). $h =4 c S N_0/\omega$.
\newline \hspace*{7mm} Note that the energy $E$ corresponds to
$\hbar \omega$ [13-14,19]. This figure is a schematic type.
\newline \hspace*{7mm} (cf. the presentation : Fig. 5 (a) for that
used in [19].) All units are dimensionless. \newline \hspace*{7mm}
As $B$ increases, the minimum (for the localization length) state
$h_{min}$ decreases, i.e., \newline \hspace*{7mm} the temperature
becomes much more lower (as the mean free path is rather large).
\end{figure}

\newpage

\psfig{file=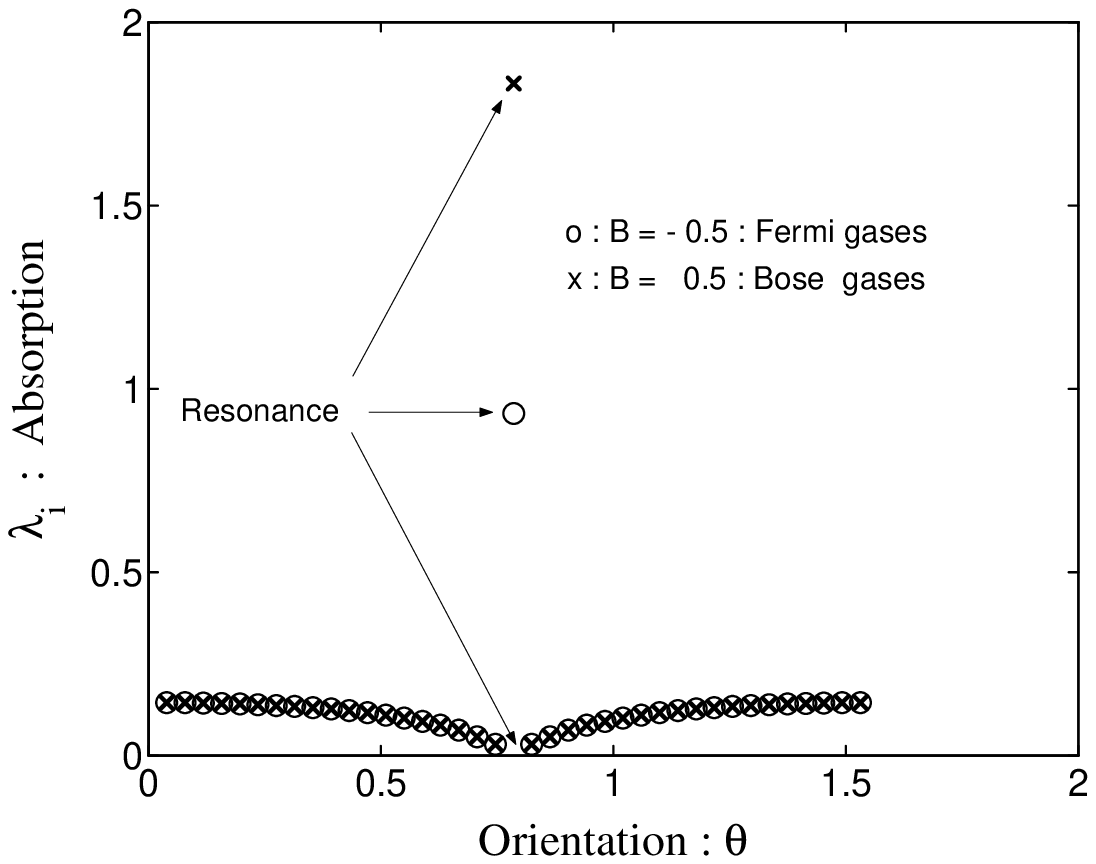,bbllx=0.2cm,bblly=11.2cm,bburx=14cm,bbury=24cm,rheight=10.8cm,rwidth=10.8cm,clip=}

\begin{figure}[h]
\hspace*{4mm} Fig. 7 \hspace*{3mm} Variations of (max.)
$\lambda_i$ w.r.t. the disorder or free orientation : $\theta$ for
\newline
\hspace*{4mm} B= $\pm 0.5$. The sudden jump at $\theta=\pi/4$
implies the resonant transition in Bose fluids \newline
\hspace*{4mm} may also occur for Fermi fluids (in bound pairs).
$\lambda_i$ for $B=0.5$ (Bose gases) \newline \hspace*{4mm} is
almost twice of that value for   $B=-0.5$ (Fermi gases) at
$\theta=\pi/4$.
\newline \hspace*{4mm}
\end{figure}

\end{document}